\documentclass[twocolumn]{aastex63}

\shorttitle{Bright [C\,{\sc ii}] emitting SMGs at $z\sim$4.6}
\shortauthors{Mitsuhashi et al.}

\graphicspath{{./}{figure2/}}
\renewcommand\baselinestretch{0.92}
\usepackage{amsmath}

\usepackage{ulem} 
\def\blue#1 {{\textcolor{blue}{#1}}\ }
\begin{document}

\title{FIR-luminous [C\,{\sc ii}] emitters in the ALMA-SCUBA-2 COSMOS survey (AS2COSMOS):\\ The nature of submillimeter galaxies in a 10 comoving Mpc-scale structure at $\emph{z}\sim4.6$}


\author{I. Mitsuhashi}
\affiliation{Department of Astronomy, The University of Tokyo, 7-3-1 Hongo, Bunkyo, Tokyo 113-0033, Japan}
\affiliation{National Astronomical Observatory of Japan, 2-21-1 Osawa, Mitaka, Tokyo 181-8588, Japan}

\author{Y. Matsuda}
\affiliation{National Astronomical Observatory of Japan, 2-21-1 Osawa, Mitaka, Tokyo 181-8588, Japan}
\affiliation{Department of Astronomy, School of Science, SOKENDAI (The Graduate University for Advanced Studies), Osawa, Mitaka, Tokyo 181-8588, Japan}

\author{Ian Smail}
\affiliation{Centre for Extragalactic Astronomy, Department of Physics, Durham University, Durham, DH1 3LE, UK}

\author{N. H. Hayatsu}
\affiliation{National Astronomical Observatory of Japan, 2-21-1 Osawa, Mitaka, Tokyo 181-8588, Japan}

\author{J. M. Simpson}
\affiliation{Centre for Extragalactic Astronomy, Department of Physics, Durham University, Durham, DH1 3LE, UK}
\affiliation{Academia Sinica Institute of Astronomy and Astrophysics, No. 1, Sec. 4, Roosevelt Rd., Taipei 10617, Taiwan}
\affiliation{National Astronomical Observatory of Japan, 2-21-1 Osawa, Mitaka, Tokyo 181-8588, Japan}

\author{A. M. Swinbank}
\affiliation{Centre for Extragalactic Astronomy, Department of Physics, Durham University, Durham, DH1 3LE, UK}

\author{H. Umehata}
\affiliation{RIKEN Clusterfor Pioneering Research, 2-1 Hirosawa, Wako, Saitama 351-0198, Japan}
\affiliation{Institute of Astronomy, Graduate School of Science, The University of 
Tokyo, 2-21-1 Osawa, Mitaka, Tokyo 181-0015, Japan}

\author{U. Dudzevi{\v{c}}i{\={u}}t{\.{e}}}
\affiliation{Centre for Extragalactic Astronomy, Department of Physics, Durham University, Durham, DH1 3LE, UK}

\author{J. E. Birkin}
\affiliation{Centre for Extragalactic Astronomy, Department of Physics, Durham University, Durham, DH1 3LE, UK}

\author{S. Ikarashi}
\affiliation{Centre for Extragalactic Astronomy, Department of Physics, Durham University, Durham, DH1 3LE, UK}

\author{Chian-Chou Chen}
\affiliation{Academia Sinica Institute of Astronomy and Astrophysics, No. 1, Sec. 4, Roosevelt Rd., Taipei 10617, Taiwan}

\author{K. Tadaki}
\affiliation{National Astronomical Observatory of Japan, 2-21-1 Osawa, Mitaka, Tokyo 181-8588, Japan}

\author{H. Yajima}
\affiliation{Center for Computational Physics, University of Tsukuba, Tsukuba, Ibaraki 305, Japan}

\author{Y. Harikane}
\affiliation{Department of Physics and Astronomy, University College London, Gower Street, London WC1E 6BT, UK}
\affiliation{National Astronomical Observatory of Japan, 2-21-1 Osawa, Mitaka, Tokyo 181-8588, Japan}

\author{H. Inami}
\affiliation{Hiroshima Astrophysical Science Center, Hiroshima University, 1-3-1 Kagamiyama, Higashi-Hiroshima, Hiroshima 739-8526, Japan}

\author{S. C. Chapman}
\affiliation{Department of Physics and Atmospheric Science, Dalhousie University, Halifax, NS B3H 3J5 Canada}

\author{B. Hatsukade}
\affiliation{Institute of Astronomy, Graduate School of Science, The University of 
Tokyo, 2-21-1 Osawa, Mitaka, Tokyo 181-0015, Japan}

\author{D. Iono}
\affiliation{National Astronomical Observatory of Japan, 2-21-1 Osawa, Mitaka, Tokyo 181-8588, Japan}
\affiliation{Department of Astronomy, School of Science, SOKENDAI (The Graduate University for Advanced Studies), Osawa, Mitaka, Tokyo 181-8588, Japan}

\author{A. Bunker}
\affiliation{Department of Physics, University of Oxford, Denys Wilkinson Building, Keble Road, OX1 3RH, UK}

\author{Y. Ao}
\affiliation{Purple Mountain Observatory \& Key Laboratory for Radio Astronomy, Chinese Academy of Sciences, 8 Yuanhua Road, Nanjing 210034, People's Republic of China}

\author{T. Saito}
\affiliation{Nishiharima Astronomical Observatory, Centre for Astronomy, University of Hyogo, 407-2 Nishigaichi, Sayo, Sayo-gun,
679-5313 Hyogo, Japan}

\author{J. Ueda}
\affiliation{National Astronomical Observatory of Japan, 2-21-1 Osawa, Mitaka, Tokyo 181-8588, Japan}

\author{S. Sakamoto}
\affiliation{Department of Astronomy, The University of Tokyo, 7-3-1 Hongo, Bunkyo, Tokyo 113-0033, Japan}
\affiliation{National Astronomical Observatory of Japan, 2-21-1 Osawa, Mitaka, Tokyo 181-8588, Japan}

\begin{abstract}
We report the discovery of a 10\,comoving\,Mpc-scale structure traced by massive submillimeter galaxies (SMGs) at $z\sim4.6$. These galaxies are selected from an emission line search of ALMA Band~7 observations targeting 184 luminous submillimeter sources ($S_{850\mu{\rm m}}\,\geq\,6.2\,\mathrm{mJy}$) across $1.6~\mathrm{degrees}^2$ in the COSMOS field. We identify four [C\,{\sc ii}] emitting SMGs and two probable [C\,{\sc ii}] emitting SMG candidates at $z=4.60$--$4.64$ with velocity-integrated signal-to-noise ratio of SNR\,$>8$. Four of the six emitters are near-infrared blank SMGs. After excluding one SMG whose emission line is falling at the edge of the spectral window, all galaxies show clear velocity gradients along the major axes that are consistent with rotating gas disks. The estimated rotation velocities of the disks are 330--550\,km\,s$^{-1}$ and the inferred host dark-matter halo masses are $\sim2$--$8\times\,10^{12}$\,M$_{\odot}$. From their estimated halo masses and [C\,{\sc ii}] luminosity function, we suggest that these galaxies have a high (50--100\%) duty cycle and high ($\sim0.1$) baryon conversion efficiency (SFR relative to baryon accretion rate), and that they contribute $\simeq2 \%$ to the total star-formation rate density at $z=4.6$. These SMGs are concentrated within just 0.3\% of the full survey volume, suggesting they are strongly clustered. The extent of this structure and the individual halo masses suggest that these SMGs will likely evolve into members of a $\sim10^{15}$\,M$_{\odot}$ cluster at $z=0$. This survey reveals synchronized dusty starburst in massive halos at $z>4$, which could be driven by mergers or fed by smooth gas accretion.
\end{abstract}


\section{Introduction}\label{sec:intro}

The connection between environment and galaxy formation is likely to be critical to understand the variety seen in the galaxy population in the local Universe \citep{1980ApJ...236..351D}. Studies of the star-formation activity of galaxies in overdense regions at high redshifts, which are expected to evolve into clusters by the present day, are a powerful tool to investigate the processes involved in galaxy formation  \citep[e.g.,][]{2001ApJ...562L...9K,2013MNRAS.434..423K}. For example, \citet{2014ApJ...790L..32R} predicts that galaxies grow predominantly by smooth gas accretion from cosmological filaments in high-redshift overdense regions. Observationally testing such claims is thus a key goal of galaxy evolution studies \citep[e.g.,][]{2019Sci...366...97U}.

Submillimeter galaxies (SMGs) are dusty strongly-star-forming galaxies at high redshift \citep[e.g.,][]{1997ApJ...490L...5S,1998Natur.394..248B,1998Natur.394..241H,2010MNRAS.405.2260S,2012MNRAS.420..957Y}, which are believed  to be tracers of massive dark-matter halos in the early Universe \citep{1999MNRAS.302..632B,2012MNRAS.421..284H,2016ApJ...820...82C,2017MNRAS.464.1380W,2019ApJ...886...48A,2020MNRAS.494.3828D}. As such SMGs are potential sign-posts to identify large-scale structures \citep{2009Natur.459...61T,2014MNRAS.440.3462U,2015ApJ...815L...8U,2019Sci...366...97U,2014PhR...541...45C,2016ApJ...824...36C,2018Natur.556..469M,2018ApJ...856...72O, 2019MNRAS.486.3047C,2020MNRAS.495.3124H,2020ApJ...889...98M} and also  provide insights into the formation processes of massive galaxies.  
The majority of SMGs are found at $z=1$--4 and they are thought to be powered by gas-rich mergers or rapid gas accretion from the cosmic web \citep[e.g.,][]{2008ApJ...680..246T,2019Sci...366...97U,2019MNRAS.488.2440M}. Recently, high sensitivity interferometric observations have revealed that  SMGs, including the higher-redshift
examples at $z>4$, tend to have rotating gas disks, which may indicate that either fast gas settling in interactions, or smooth gas accretion, make an important contribution to fuel  high-redshift starbursts \citep{2012ApJ...760...11H,2014A&A...565A..59D,2018Natur.560..613T}. To understand the formation process of $z>4$ SMGs in the cosmological context, we need to study their statistical properties based on wide-field spectroscopic survey of $z>4$ SMGs. 

The $^2$P$_{3/2}$--$^2$P$_{1/2}$ fine structure line of C$^+$ at 157.74\,$\mu$m (hereafter [C\,{\sc ii}]) is one of the brightest far-infrared lines \citep[e.g.,][]{2008ApJS..178..280B,2011MNRAS.414L..95S}. The [C\,{\sc ii}] line is suitable for spectroscopic identifications and studies of gas dynamics of $z>4$ SMGs at submillimeter wavelengths \citep[e.g.,][]{2006ApJ...645L..97I,2014A&A...565A..59D,2017ApJ...850..180J}. For instance, the ALMA-LABOCA Extended $\textit{Chandra}$ Deep Field-South Survey \citep[ALESS,][]{2013ApJ...768...91H} and ALMA-SCUBA-2 survey of the UDS field \citep[AS2UDS,][]{2019MNRAS.487.4648S} serendipitiously detected [C\,{\sc ii}] emitting SMGs at $z=4.4$--4.6 from their ALMA Band~7 snapshot observations of large samples of submillimeter sources \citep{2012MNRAS.427.1066S, 2018ApJ...861..100C}. 

\indent In this paper, we report a search for [C\,{\sc ii}]-emitting SMGs in the Cosmic Evolution Survey (COSMOS) field, based on ALMA Band~7 snapshot observations. We take advantage of the [C\,{\sc ii}] line to investigate the star-formation activity in the SMGs and the environments they reside in. Throughout this paper, we assume a $\Lambda$CDM cosmology with $\Omega_{{\rm M}} = 0.3$, $\Omega_{\Lambda} = 0.7$ and $H_0 = 70$\,km\,s$^{-1}$\,Mpc$^{-1}$.

\section{Data}\label{sec:data}

The targets studied in this work  represent an effectively complete sample of 184 luminous ($S_{850\mu\mathrm{m}}\geq6.2$\,mJy) submillimeter sources across a sky area of $\sim1.64$ degrees$^2$, selected from the SCUBA-2 COSMOS survey \citep[S2COSMOS,][]{2019ApJ...880...43S}. Among the 184 sources, ALMA Band~7 data of 24 sources are obtained from the ALMA data archive (2013.1.00034.S 2015.1.00137.S; 2015.1.00568.S; 2015.1.01074.S; 2016.1.01604.S; 2016.1.00478.S). We observed the remaining 160 submillimeter sources in program 2016.1.00463.S and full details of the observation, reduction and analysis of the whole survey are presented in \citet{2020MNRAS.495.3409S}. The observations were performed with standard single continuum setup with a total bandwidth of 7.5~GHz centered on 344\,GHz, split into two sidebands (335.7--339.3\,GHz and 347.7--351.4\,GHz), corresponding to frequency of the redshifted [C\,{\sc ii}] emission line in sources at $z=4.40$--$4.46$ and $z=4.60$--$4.66$ respectively. We estimate that the  full survey volume for the [C\,{\sc ii}] line emission search in our study is $2.2 \times 10^6$\,cMpc$^3$. The full width half maximum (FWHM) of the ALMA primary beam at 344\,GHz is $\sim17''$, which covers the whole SCUBA-2 beam ($\sim15''$). The minimum and maximum baseline lengths were 15.0~m and 313.7~m respectively.

We calibrated the data and made dirty image cubes with Common Astronomy Software Application ({\sc casa}) version 5.6.1 \citep{2007ASPC..376..127M}. The cubes have a synthesized beam of $0.8''\times0.8''$ and a typical $1$-$\sigma$ depth of $\sim 2.4$\,mJy\,beam$^{-1}$ in a 27\,km\,s$^{-1}$ spectral channel. A more detailed description of reduction and the full catalogue of the 260 continuum sources are presented in \citet{2020MNRAS.495.3409S}.

\section{Analysis and Results}\label{sec:analysis}

%
%
\begin{figure}[b]
\epsscale{1.0}
\centering
\includegraphics[width = 8.2cm,trim = 0.0 6.0 0.0 0.0 cm]{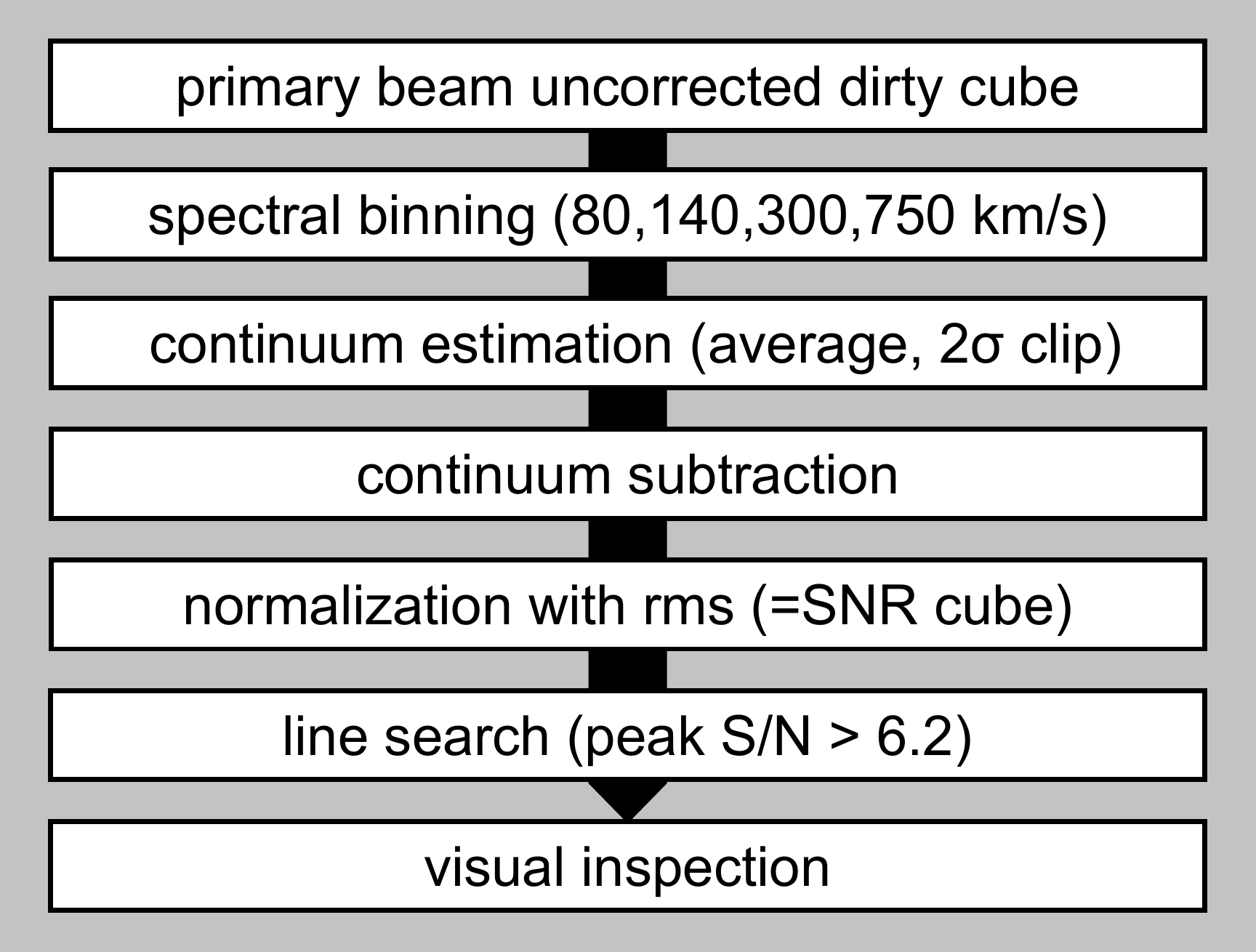}
\caption{Flowchart of our line search method.}
\label{fig:FC}
\end{figure}

%
%
\begin{figure*}[htp]
\epsscale{1.15}
\includegraphics[width = 17.8 cm,trim = 0.5 0 0 0 cm]{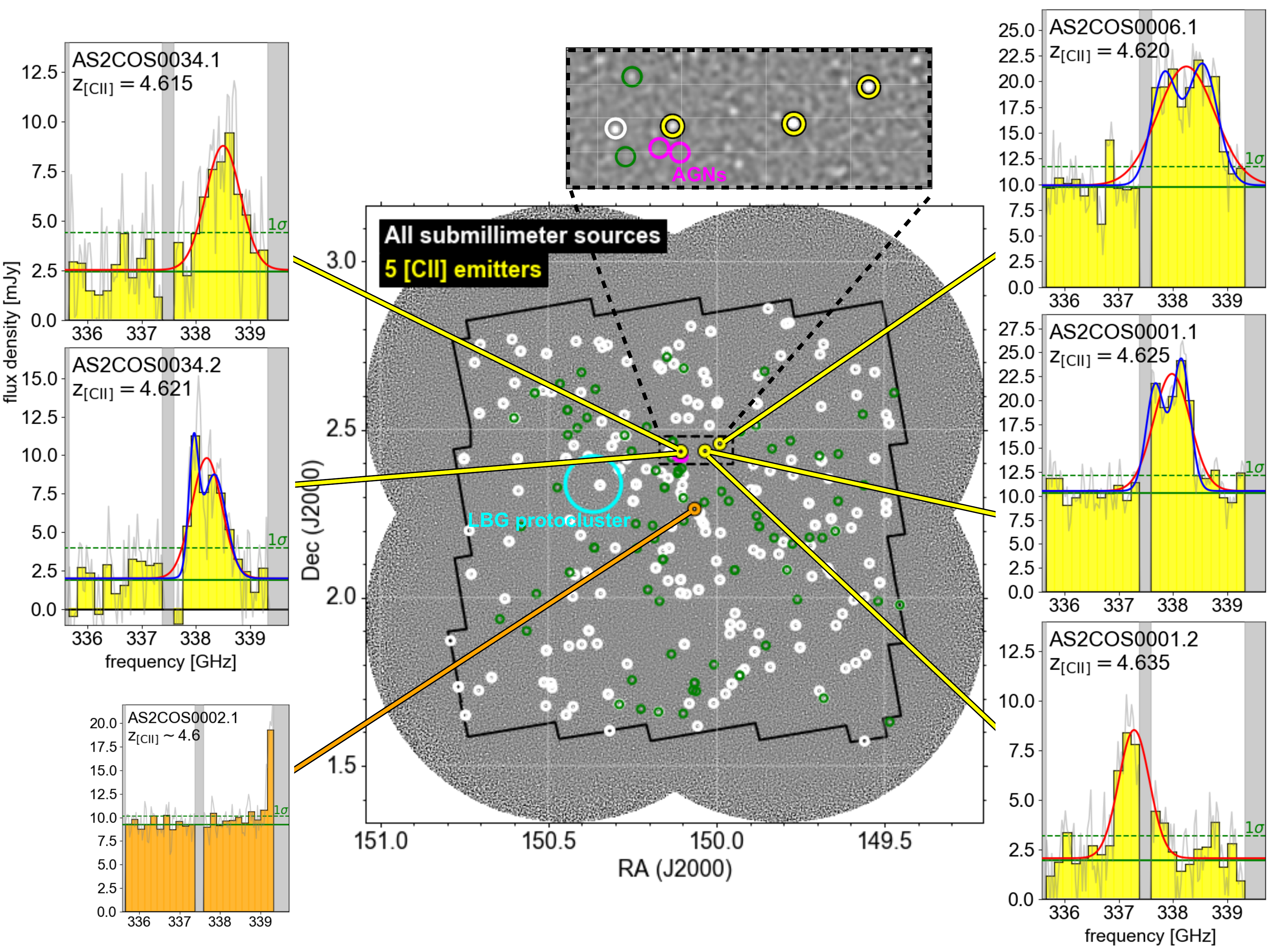}
\caption{Sky positions of five [C\,{\sc ii}]-emitting SMGs (yellow circles) and the emitter we excluded (orange circle) with the spatially integrated spectra of their emission lines.  The white circles show the positions of the parent sample of 184 SMGs from the AS2COSMOS survey \citep{2020MNRAS.495.3409S} and green circles show the positions of the sources from S2COSMOS and A$^3$COSMOS \citep{2019ApJ...886...48A,2019ApJS..244...40L} with photometric redshift of $z=4$--5. The five emitters are clustered in a small sky region ($\sim 7'$ or $16$\,comoving Mpc) and in a narrow redshift range ($\Delta z_{\text{[C\,{\sc ii}]}}=0.021$ or $\Delta v = 1100$\,km\,s$^{-1}$).  The background image is the 850-{$\mu$m} SCUBA-2 map from \citet{2019ApJ...880...43S} and the black solid line shows the survey area of 1.6 degree$^2$ corresponding to the \textit{HST}/ACS coverage. We also show two AGNs at $z=4.64$ \citep[][magenta circles]{2018ApJ...858...77H} and a protocluster at $z=4.53$--4.60 \citep[][cyan circle]{2018A&A...615A..77L}. The spectra were binned to a $\sim 135\,\mathrm{km\,s}^{-1}$ resolution. Red and blue lines show single and double Gaussian fits respectively. The dashed lines show the average 1-$\sigma$ noise of each channel. The gray shaded regions in each spectrum show the gaps in the spectral coverage. }
\label{fig:sky}
\end{figure*}

\indent We search for emission lines in signal-to-noise ratio (SNR) cubes as shown in Figure~\ref{fig:FC} \citep[see also][]{2017PASJ...69...45H}. We start from the dirty cubes without primary beam correction, which have uniform noise levels across each field. To provide sensitivity to a wide range of line widths, we  bin these data cubes to four different velocity resolutions \citep[80, 140, 300 and 750\,km\,s$^{-1}$, see][]{2018ApJ...861..100C}. 
To estimate the continuum level excluding emission lines, we average all channels for each spatial pixel in the binned data cubes using  2-$\sigma$ clipping. We iterate this procedure until the 2-$\sigma$ clipping  converges, and subtract the resulting continuum values from the data cubes. We check that the 2-$\sigma$ clipped average is more stable than a linear fit without over-subtraction. We also confirm that continuum levels estimated from a median agree with those from 2-$\sigma$ clipped average within 10\% and that the choice of the methods used in continuum estimation does not affect the following line detection. As the noise is not constant across the frequency bands, we construct the SNR cubes by dividing each channel of the binned dirty cubes by the rms of that channel measured across the full field of the cube.

In the binned, continuum-subtracted SNR cubes, we detect six line emitters with peak SNR $>6.2$. We adopt this $6.2$-$\sigma$ threshold on the grounds that no false detection appear in any of the  inverted data cubes above this significance. This means that there is no negative peak exceeding $-6.2\sigma$ in  any of the 184 SNR cubes we searched. For 80, 140, 300, 750\,km\,s$^{-1}$ binning scales, this threshold corresponds to detection limits of [{C\,{\sc ii}}] luminosity of $L_{\text{[{C\,{\sc ii}}]}}=0.5, 0.7, 1.1, 1.8\times10^9$\,L$_{\odot}$ at $z\sim4.5$, or observed frame equivalent width (EW) of EW = 0.4, 0.5, 0.8, 1.3\,$\mu$m for a source with $S_{870\mu\mathrm{m}}=6.2$\,mJy. All of the six line emitters are associated with 870-$\mu$m continuum sources. We exclude one line emitter (AS2COS0002.1) because we cannot measure the total line flux or line velocity width of the emission line as it falls at the edge of the spectral window\footnote{We confirm that AS2COS0002.1 (R.A.\ 150.106505\,deg, Dec.\ 2.26362\,deg) has a redshift of $z=4.596$ through the detection of $^{12}$CO(5--4) (Chen et al.\ in prep) and has a faint near-infrared counterpart ($K_s>24$\,ABmag).} (see Figure \ref{fig:sky}). We measure the SNR of the emission lines from velocity integrated maps between $\nu_\mathrm{obs}-0.5\times\mathrm{FWHM}$ and $\nu_\mathrm{obs}+0.5\times\mathrm{FWHM}$ at the peak pixels.  Our final sample comprises  five  emitters with lines detected at  high significance levels, integrated SNR\,$>8$. We show the sky distribution of the emitters in Figure \ref{fig:sky}. All of these emitters  fall within a 7-arcmin diameter region and a narrow frequency range (337.3--338.5\,GHz).

\begin{figure}[tp]
\vspace{10pt}
\epsscale{1.0}
\includegraphics[width = 9.0cm,trim = 20.0 0.0 0.0 30.0 cm]{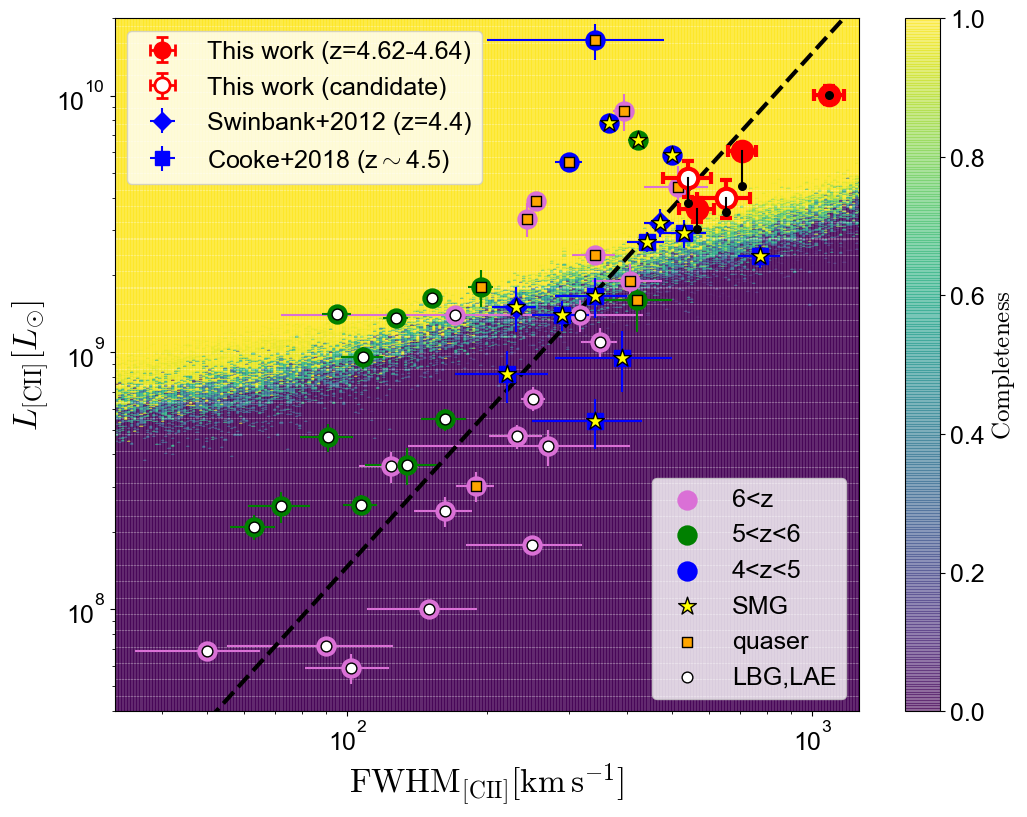}
\caption{[C\,{\sc ii}] velocity width vs line luminosity of the five emitters. Red and black circles show the values before and after correcting for lensing amplification (See Section \ref{subsec:lens}). Black dashed line shows $L_{\text{[C\,{\sc ii}]}}\propto\mathrm{FWHM}_{\text{[C\,{\sc ii}]}}^2$ relation scaled to the data points of SMGs. Background shows the completeness of our line search as a function of the input velocity width and input line luminosity. We confirm that the completeness is almost unity above $L_{\text{[C\,{\sc ii}]}}=3\times10^9\,L_{\odot}$. The literature values come from \citet{2006ApJ...645L..97I,2010A&A...519L...1W,2013ApJ...773...44W,2013ApJ...770...13W,2013ApJ...778..102O,2013A&A...559A..29C,2018ApJ...854L...7C,2014ApJ...796...84R,2015Natur.522..455C,2015MNRAS.452...54M,2015ApJ...807..180W,2015MNRAS.454.3485Y,2016ApJ...829L..11P,2016arXiv161108552M,2016ApJ...816...37V,2018Natur.553..178S,2019PASJ...71...71H,2020ApJ...896...93H}}
\label{fig:lcii_fwhm}
\end{figure}

We have checked the completeness of our line search by simulating $1.2\times10^7$ Gaussian-like emission line with the range of the FWHMs of $80$--$2000$~km~s$^{-1}$ and luminosities of $10^{7.5}$--$10^{10.1}L_{\odot}$ and injecting them in data cubes containing pure noise representative of the survey (2.4\,mJy\,beam$^{-1}$ per channel). We split FWHMs into 300 cells and luminosities into 700 cells in log space, thus there are 55 lines in each cell. From this analysis we find that the completeness is almost unity above $L_{\text{[C\,{\sc ii}]}}=3\times10^9\,L_{\odot}$, which is about $\sim10$ times higher than the detection limits for other [C\,{\sc ii}] emitting galaxies at $z>4$ (see Figure \ref{fig:lcii_fwhm}).

\begin{figure*}[htp]
\epsscale{1.15}
\centering
\includegraphics[width = 18 cm,trim = 0.0 0.0 0.0 0.0 cm]{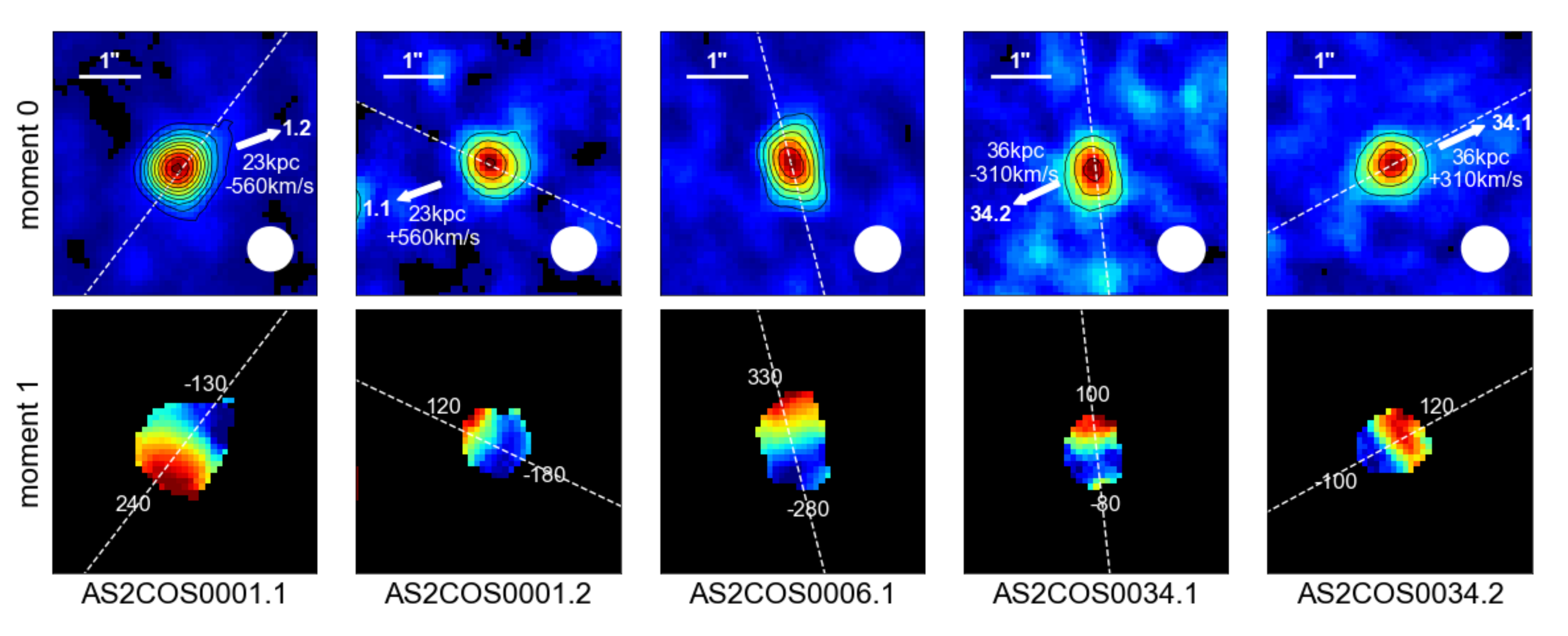}
\caption{Top: The velocity-integrated flux maps of  five detected line-emitting SMGs. The black contours show the surface brightness of the emission lines starting from $3\sigma$ with $2$-$\sigma$ intervals. Velocity offsets, projected spatial separations and directions of any companions (if they have one) are also shown. The dashed lines show major-axis position angles. Bottom: The velocity maps with $3$-$\sigma$ clipping. Typical velocities (in km\,s$^{-1}$) and major axis position angles are also denoted. All of the five sources show clear velocity gradients along the major axes that are consistent with rotating  disks.}  \label{fig:mommap}
\end{figure*}

As the SNRs of our detected lines are not high enough to require {\sc clean}ing, we measure the line properties using the dirty, primary-beam-corrected flux data cubes. We use {\sc casa/imfit} task to measure the continuum flux densities, velocity-integrated line fluxes and sizes. The observed properties are summarized in Table~\ref{tab:lineobs}. We re-calculate continuum levels excluding channels containing the emission lines and construct line-free continuum maps. The line-free continuum maps are used to measure the spatially integrated continuum flux densities with 2-D Gaussian profiles. We also construct emission line cubes by subtracting the estimated continuum values (shown as green solid lines in Figure~\ref{fig:sky}) from the cubes. We measure the velocity widths of the spatially-integrated emission lines using single Gaussian profiles (see Figure~\ref{fig:sky}). All of the emission lines have large velocity widths: $\gtrsim 500\,{\mathrm{km\,s}^{-1}}$. Figure~\ref{fig:mommap} shows the velocity-integrated flux maps (moment~0) and the velocity maps (moment~1) of the five emitters. The spatially-integrated line fluxes and beam-deconvolved major/minor axes are measured by using 2-D Gaussian profiles for the line emitting regions detected above $>3\sigma$ on the moment~0 maps. We correct for missing line flux falling into the gaps of the spectral windows by interpolating the emission lines with single Gaussian fits. The estimated missing line fluxes are $\sim10\%$ for AS2COS0001.1, $\sim 25\%$ for AS2COS0001.2 and negligible (less than $5\%$) for the remaining three sources. All the five emitters show clear velocity gradients on the moment~1 maps. Since the directions of these velocity gradients match the major axis position angles, they are likely to be rotating gas disks. \citet{2020MNRAS.495.3409S} and \citet{2020ApJ...890..171J} have already confirmed that AS2COS0001.1/1.2 show clear rotating features with higher angular resolution ALMA data.

%
%

\subsection{Line identification}\label{subsec:cii}

%
%
\begin{deluxetable*}{cccccc}
\tablenum{1}
\tablecaption{Properties of the five emission-line-detected SMGs \label{tab:lineobs}}
\tablewidth{0pt}
\tablehead{\colhead{ID$^{\scalebox{1.1}{\it{a}}}$} & \colhead{\hspace{5pt} AS2COS0001.1}\hspace{5pt} & \colhead{\hspace{5pt}AS2COS0001.2}\hspace{5pt} & \colhead{\hspace{5pt}AS2COS0006.1}\hspace{5pt} & \colhead{\hspace{5pt}AS2COS00034.1}\hspace{5pt} & \colhead{\hspace{5pt}AS2COS0034.2}\hspace{5pt}
}
\startdata
R.A.(deg)\tablenotemark{\small a} & 150.03350 & 150.03268 & 149.98871 & 150.10446 & 150.10572\\
Dec.(deg)\tablenotemark{\small a} & 2.43675 & 2.43701 & 2.45850 & 2.43537 & 2.43481\\
$\mu S_{870\mu\mathrm{m}}$(mJy)\tablenotemark{\small b} & $12.2\pm0.4$ & $3.2\pm0.2$ & $12.5\pm0.5$ & $4.1\pm0.3$ & $3.7\pm0.2$\\
$\mu S_{3\mathrm{GHz}}$($\mu$Jy)\tablenotemark{\small c} & $15.0\pm2.4$ & $28.8\pm2.7$ & $13.2\pm3.3$ & $13.0\pm2.4$ & $3.9\pm2.2$\\
$\nu_{{\rm obs}}$(GHz) & 337.90 & 337.27 & 338.17 & 338.49 & 338.14\\
$z_{\text{[C\,{\sc ii}]}}$ & $4.625\pm0.001$ & $4.635\pm0.001$ & $4.620\pm0.001$ & $4.615\pm0.001$ & $4.621\pm0.001$\\
FWHM$_{\text{[C\,{\sc ii}]}}$(km s$^{-1}$) & $710\pm50$ & $560\pm50$ & $1090\pm80$ & $650\pm80$ & $540\pm60$\\
$Sdv_{\text{[{C\,{\sc ii}}]}}$(Jy km s$^{-1}$)\tablenotemark{\small d} & $7.0^{+0.8}_{-1.3}$ & $4.7^{+0.6}_{-0.6}$ & $15.9\pm1.3$ & $5.5^{+1.0}_{-1.1}$ & $6.0^{+1.1}_{-1.7}$\\
$L_{\text{[{C\,{\sc ii}}]}}$($10^{9}L_{\odot}$) & $4.4^{+0.5}_{-0.8}$ & $3.0^{+0.4}_{-0.4}$ & $10.1\pm0.8$ & $3.5^{+0.6}_{-0.7}$ & $3.8^{+0.7}_{-1.1}$\\
SFR$_{\text{[{C\,{\sc ii}}]}}(M_{\odot}$ yr$^{-1}$) & $390^{+350}_{-180}$ & $260^{+240}_{-120}$ & $880^{+790}_{-420}$ & $310^{+280}_{-150}$ & $330^{+300}_{-160}$\\
SNR\tablenotemark{\small e} & 15.3 & 10.0 & 17.4 & 8.0 & 10.7\\
EW($\mu$m)\tablenotemark{\small f} & $2.3\pm0.2$ & $5.2\pm0.7$ & $3.7\pm0.3$ & $4.5\pm0.8$ & $5.9\pm1.0$\\
size$_{\text{[{C\,{\sc ii}}]}}(~''\times~''$)\tablenotemark{\small g} & $0.6^{\pm0.1}\times0.4^{\pm0.1}$ & $0.6^{\pm0.1}\times0.5^{\pm0.2}$ & $0.9^{\pm0.1}\times0.2^{\pm0.1}$ & $0.8^{\pm0.2}\times0.1^{\pm0.3}$ & $0.6^{\pm0.3}\times0.4^{\pm0.3}$\\
$i$(deg) & $47\pm9$ & $34^{+28}_{-34}$ & $77\pm8$ & $80^{+10}_{-21}$ & $52^{+34}_{-52}$\\
$V_\mathrm{rot}$(km s$^{-1}$) & $470\pm80$ & $490_{-180}$\tablenotemark{\small h} & $550\pm40$ & $330^{+50}_{-10}$ & $340_{-70}$\tablenotemark{\small h}\\
$M_h^{\ \mathrm{ind}}$($10^{12} $\,M$_{\odot}$) & $5.1\pm1.4$ & $5.7_{-3.6}$\tablenotemark{\small h} & $8.0\pm1.1$ & $1.7^{+0.4}_{-0.1}$ & $1.9_{-0.7}$\tablenotemark{\small h} \\
$M_h^{\ \mathrm{NFW}}$($10^{12}$\,M$_{\odot}$)\tablenotemark{\small i} & \multicolumn{2}{c}{$6.4^{+2.3}_{-1.8}$} & - & \multicolumn{2}{c}{$0.8^{+0.6}_{-0.4}$}\\
$M_\mathrm{dyn}$($10^{11}$\,M$_{\odot}$) & $1.7\pm0.3$ & $1.9_{-0.9}\tablenotemark{\small h}$ & $2.3\pm0.3$ & $0.8^{+0.2}_{-0.1}$ & $0.9_{-0.3}\tablenotemark{\small h}$\\
SFR$_{870\mu\mathrm{m}}(M_{\odot}$ yr$^{-1}$)\tablenotemark{\small j} & $390^{+80}_{-10}$ & $230^{+30}_{-10}$ & $450^{+10}_{-10}$ & $260^{+30}_{-10}$ & $240^{+40}_{-10}$\\
$\mu$\tablenotemark{\small k} & $1.4^{+0.2}_{-0.1}$ & $1.2^{+0.1}_{-0.1}$ & - & $1.1^{+0.1}_{-0.1}$ & $1.3^{+0.3}_{-0.1}$
\enddata

\tablecomments{
\vspace{-6pt}\tablenotetext{\small a}{Source IDs and coordinates come from the full AS2COSMOS catalog presented in \citet{2020MNRAS.495.3409S}.}
\vspace{-6pt}\tablenotetext{\small b}{total continuum flux densities excluding line emission measured with the {\sc CASA/imfit} task.}
\vspace{-6pt}\tablenotetext{\small c}{radio continuum fluxes from \citet{2017AandA...602A...1S}.}
\vspace{-6pt}\tablenotetext{\small d}{total emission line flux measured with the {\sc CASA/imfit} task.}
\vspace{-6pt}\tablenotetext{\small e}{velocity-integrated peak emission line signal to noise ratio within the FWHM.}
\vspace{-6pt}\tablenotetext{\small f}{observed-frame line equivalent widths.}
\vspace{-6pt}\tablenotetext{\small g}{beam deconvolved major and minor axes measured with the {\sc CASA/imfit} task.}
\vspace{-6pt}\tablenotetext{\small h}{upper limits cannot be constrained from uncertainty of the inclinations.}
\vspace{-6pt}\tablenotetext{\small i}{common halo masses assuming NFW dark matter profile.}
\vspace{-6pt}\tablenotetext{\small j}{star-formation rate estimated from $S_{870\mu\mathrm{m}}$ and conversion in \citet{2020MNRAS.494.3828D}.}
\vspace{-6pt}\tablenotetext{\small k}{gravitational magnification factors.}
}
\vspace{-22pt}
\end{deluxetable*}

\indent Our five line-emitting SMGs have also been detected with AzTEC on the James Clerk Maxwell Telescope (JCMT), with AS2COS0001 and AS2COS0006 corresponding to AzTEC2 and AzTEC9 \citep{2008MNRAS.385.2225S,2009ApJ...704..803Y}. Later, AS2COS0001, AS2COS0006 and AS2COS0034 were also identified as AzTEC-C3, AzTEC-C14 and AzTEC-C30 using AzTEC on the Atacama Submillimeter Telescope Experiment \citep[ASTE,][]{2011MNRAS.415.3831A}. These SMGs are also detected in the {\it Herschel}/HerMES survey \citep{2012MNRAS.424.1614O} and an earlier SCUBA-2 850\,$\mu$m survey \citep{2013MNRAS.436.1919C}. While more recently, AS2COS0001.1/1.2, AS2COS0006.1 and AS2COS0034.1/34.2 were identified in ALMA~Band6 as AzTEC-C3a/C3b, AzTEC-C14 and AzTEC-C30a/C30b \citep{2017A&A...608A..15B}.

Among the five line emitters, three have been already confirmed to be at $z\sim4.6$. Recently, \citet{2020ApJ...890..171J} observed the two line emitters (AS2COS0001.1/1.2) with NOrthern Extended Millimeter Array (NOEMA) Band~1 and detected $^{12}$CO(5--4) lines at $z=4.63$ \citep[see also][]{2020MNRAS.495.3409S}. \citet{2020arXiv200903341B} also observed AS2COS0006.1 as a part of large CO surveys of luminous SMGs and detected $^{12}$CO(5--4) and [C\,{\sc i}] at $z=4.62$.

%
%

%
%
\begin{figure*}[htp]
\epsscale{1.15}
\centering
\includegraphics[width = 16 cm,trim = 0.0 30.0 0 0.0 cm]{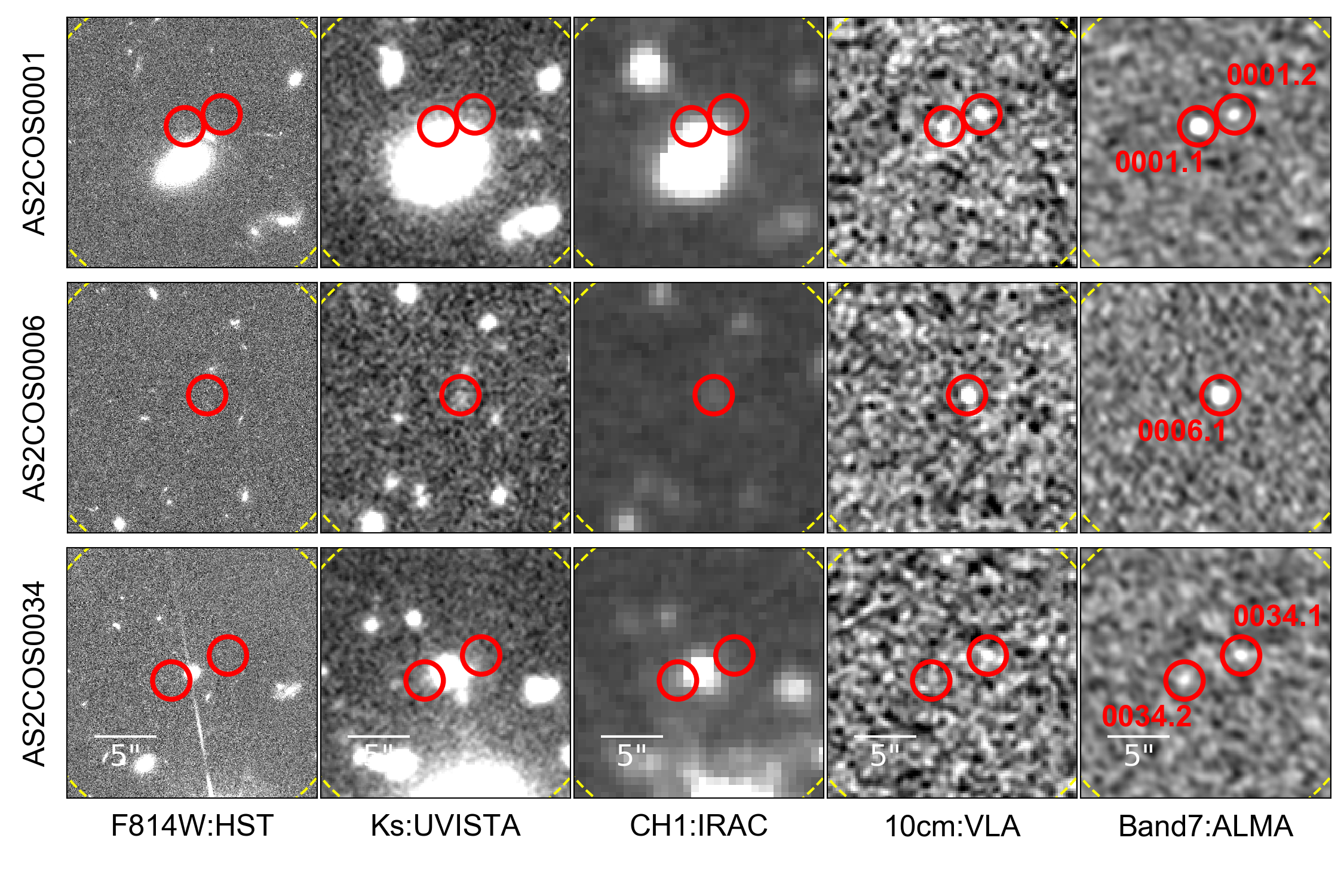}
\caption{Multi-wavelength images of the five  line-emitting SMGs. The small red circles show the positions of ALMA Band~7 continuum sources (Table \ref{tab:lineobs}). The yellow dashed circles indicates the ALMA field of view.  AS2COS0034.1/34.2 do not have detected counterparts in the optical or near-infrared bands (similar to AS2COS0001.1/1.2), suggesting that they are also likely to be at $z>4$. In the fields of AS2COS0001 and AS2COS0034, there are foreground galaxies near the SMGs, which could modestly magnify the sources due to gravitational lensing (see Section \ref{subsec:lens}). 
\label{fig:mwlimage}}
\end{figure*}

%
%
\begin{figure}[tp]
\vspace{10pt}
\epsscale{1.0}
\includegraphics[width = 8.2cm,trim = 30.0 10.0 0.0 0.0 cm]{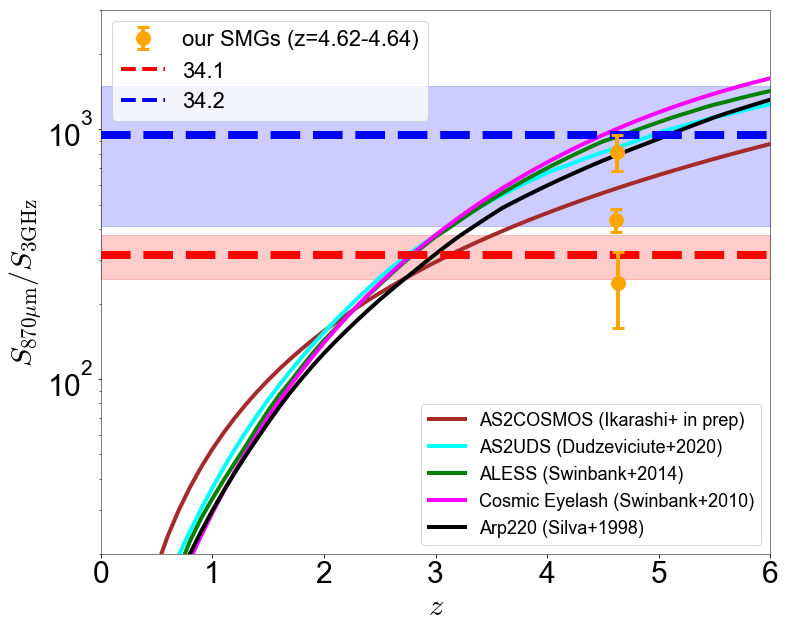}
\caption{The $S_{870\mu\mathrm{m}}/S_{3\mathrm{GHz}}$ flux ratio as a function of redshift. The colored curves show the predicted ratios based on the SEDs of Arp220, Cosmic Eyelash and various composite SMGs \citep[][Ikarashi et al. in prep]{1998ApJ...509..103S,2010Natur.464..733S,2014MNRAS.438.1267S,2020MNRAS.494.3828D}. The flux ratios of AS2COS0034.1/34.2 suggest that they are likely to lie at $z=3$--5 and are comparable with the ratios of the confirmed [{C\,{\sc ii}}]-selected SMGs at $z=4.62$--$4.64$.}
\label{fig:ratio}
\end{figure}

Their multi-wavelength properties,  line equivalent widths and close proximity suggest that the remaining two sources (AS2COS0034.1/34.2) are also likely to be [C\,{\sc ii}] emitters at $z=4.62$. Figure~\ref{fig:mwlimage} shows multi-wavelength images of the five emitters \citep{2007ApJS..172..196K,2007ApJS..172...86S,2012A&A...544A.156M,2017AandA...602A...1S}. The confirmed [C\,{\sc ii}] emitters are near-infrared blank or faint SMGs ($K_s>24$\,ABmag). AS2COS0034.1/34.2 do not have any clear optical/near-infrared counterparts, which is consistent with them lying at $z\gtrsim4$ \citep[e.g.,][]{2014ApJ...788..125S,2019Natur.572..211W,2020MNRAS.494.3828D,2020A&A...640L...8U,2020arXiv201002250S}. We use the $S_{870\mu\mathrm{m}}/S_{3\mathrm{GHz}}$ flux ratio as a crude redshift indicator \citep[e.g.,][]{1999ApJ...513L..13C,2000AJ....119.2092B,2000ApJ...528..612S}. AS2COS0034.1/34.2 have $S_{870\mu\mathrm{m}}/S_{3\mathrm{GHz}}$ flux ratios of $320\pm60$ and $950\pm540$ respectively. By comparing with the typical SEDs of SMGs \citep[][Ikarashi et al. in prep]{1998ApJ...509..103S,2010Natur.464..733S,2014MNRAS.438.1267S,2020MNRAS.494.3828D}, the flux ratios suggest that they are likely to be at $z=3$--$5$ (see Figure \ref{fig:ratio}). In addition, these values are consistent with those of our three confirmed [C\,{\sc ii}]-emitters at $z=4.62$--$4.64$. Thus AS2COS0034.1/34.2 are unlikely to be low- or mid-$J$ CO line emitters at $z\ll3.0$. Moreover, the emission lines of AS2COS0034.1/34.2 have observed-frame equivalent widths of $\mathrm{EW}=4.5\pm0.8$ and $5.9\pm1.0$\,$\mu$m, which are similar to the [C\,{\sc ii}] EW of the $z \sim 4.4$ SMGs (1.8 and 4.9\,$\mu$m) from \citet{2012MNRAS.427.1066S} and our three confirmed [C\,{\sc ii}]-emitters at $z=4.62$--$4.64$ (2.3--5.3\,$\mu$m). These observed-frame EWs are several times larger than high-$J$ CO lines at intermediate redshifts \citep[e.g.,][]{2017PASJ...69...45H} and three times smaller than [O\,{\sc iii}] at $z\sim9$ \citep[e.g.,][]{2019ApJ...874...27T}. Thus AS2COS0034.1/34.2 are unlikely to be high-$J$ CO line or [O\,{\sc iii}] emitters. Therefore we treat these two sources as [C\,{\sc ii}] emitter candidates at $z\sim4.6$.

\subsection{Gravitational lensing}\label{subsec:lens}

There is a $z=0.73$ rich cluster 5-arcmin north-east from AS2COS0006.1 \citep{2007ApJS..172..254G}, which could amplify the emission from the SMGs either due to cluster-scale  gravitational lensing \citep{2011MNRAS.415.3831A}, or more realistically given the angular separation to the cluster center, by increasing the projected density of foreground galaxies, which could act as galaxy-scale lenses \citep[e.g.,][]{2005ApJ...631..121S}. As shown in Figure~\ref{fig:mwlimage}, there are foreground galaxies near the SMGs in the fields of AS2COS0001.1/1.2 and AS2COS0034.2/34.2 which could also amplify the emission from the SMGs by galaxy-galaxy gravitational lensing (the galaxy close to AS2COS0034.1/34.2 is
indeed at $z=0.73$). To assess the relative importance of these various
lensing structures we estimate the  amplification factor, $\mu$, by using a simple singular isothermal sphere model.

The foreground $z=0.73$ cluster has an estimated  mass within $R_{500}$ (84 arcsec) of $M\sim 1.7 \times10^{14}$\,M$_{\odot}$ \citep{2007ApJS..172..254G}. 
An isothermal model would suggest a virial velocity of $V_{\mathrm{vir}}=1100$\,km\,s$^{-1}$ within a virial radius of $r_{\mathrm{vir}}=1200$\,kpc. 
We extrapolate the isothermal distribution to the source separations and convert from the rotational velocity to velocity dispersion with the relationship of $\sigma^2=V_{\mathrm{vir}}^2/2$. The resulting predicted amplifications for AS2COS0001, AS2COS0006 and AS2COS0034 are $\mu=1.02\pm0.01$, $1.04\pm0.01$ and $1.01\pm0.01$, which are less than 5\% and negligible.

A foreground galaxy at $z=0.33$ is located near AS2COS0001.1/1.2 \citep{2008ApJS..176...19F}. This galaxy has an estimated stellar mass of $M_{\star}=10^{11.0}$\,M$_{\odot}$ and is classified as a quiescent galaxy \citep{2016ApJS..224...24L}. We convert the stellar mass to a velocity dispersion of $170\pm30$\,km\,s$^{-1}$  using the Faber-Jackson relation in \citet{2006MNRAS.370.1106G}. The predicted amplifications for AS2COS0001.1/1.2 are then $\mu=1.4^{+0.2}_{-0.1}$ and $1.2^{+0.1}_{-0.1}$ respectively. We confirm that these amplification factors are almost comparable with the result in the previous work \citep[][1.5$\times$ and 1.35$\times$ respectively]{2020ApJ...890..171J}.

Similarly, a foreground galaxy at $z=0.73$ is also located between AS2COS0034.1/34.2 \citep{2007ApJS..172...70L}. We note that different velocity peaks and different EW of the emission lines (see Figure~\ref{fig:sky} and Table~\ref{tab:lineobs}) suggest that these two sources are not strongly-lensed multiple images of a single galaxy. The foreground galaxy has an estimated  stellar mass of $M_{\star}=10^{10.6}$\,M$_{\odot}$ and is classified as a star-forming galaxy \citep{2016ApJS..224...24L}. We derive a rotation velocity of $V_{\mathrm{rot}}=200^{+70}_{-50}$\,km\,s$^{-1}$ from the Tully-Fisher relation in \citet{2016A&A...594A..77D}.  We then convert from the rotational velocity to velocity dispersion with the relationship of $\sigma^2=V_{\mathrm{rot}}^2/2$. The resulting predicted amplifications for AS2COS0034.1/34.2 are $\mu=1.1^{+0.1}_{-0.1}$ and $1.3^{+0.3}_{-0.1}$ respectively.

The influence of galaxy-galaxy lensing seems dominant for AS2COS0001.1/1.2 and AS2COS0034.1/34.2, rather than the effect of the foreground $z=0.73$ cluster. Although we have confirmed that the expected lens amplifications due to foreground galaxies are relatively modest, we correct for these magnification factors throughout the rest of the paper.

\subsection{Spatial distribution and environment}\label{subsec:dist}

%
%
\begin{figure}[htp]
\epsscale{1.0}
\includegraphics[width = 8.2cm,trim = 30.0 10.0 0.0 0.0 cm]{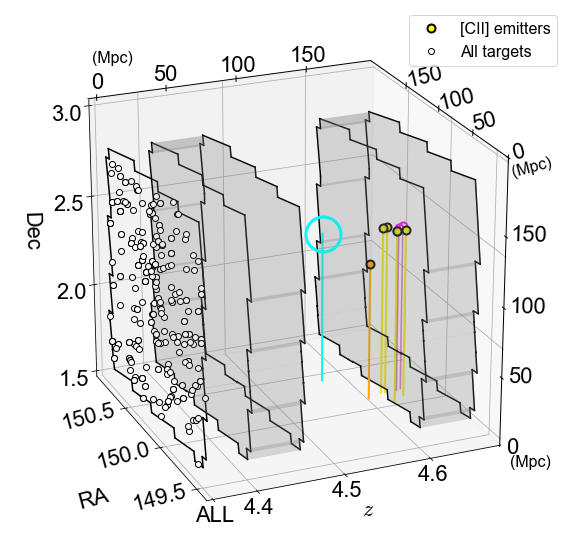}
\caption{3-D map of [C\,{\sc ii}] emitters (and candidates) in the full survey volume. Symbols are as in Figure \ref{fig:sky}. We see that the [C\,{\sc ii}] emitters are both spatially concentrated on the sky (as shown in )Figure~\ref{fig:sky}), but also in 3-D space.}
\label{fig:3Dplot}
\end{figure}

Figure~\ref{fig:3Dplot} shows the 3-D map of our five [C\,{\sc ii}] emitting SMGs (and candidates). We see that these SMGs are concentrated within a $7.0\times1.4$-arcmin region on the sky ($16\times3$\,cMpc at $z=4.6$) and a narrow-redshift range of $\Delta z=0.021$ (12\,cMpc at $z=4.6$), which correspond to just $\simeq0.3$\% of the full survey volume (gray shaded regions in Figure \ref{fig:3Dplot}, see Section~\ref{sec:data}). The
SMG distribution appears to link to the surrounding larger scale structure (see Figure~\ref{fig:sky} and \ref{fig:3Dplot}). AS2COS0002.1 at $z=4.596$ is located 10-arcmin (20\,cMpc in projection) south from the SMG structure. There are two AGNs at $z=4.64$ very close to the SMG structure \citep{2018ApJ...858...77H}. There is also a protocluster of LBGs at $z=4.53$--$4.60$ \citep{2018A&A...615A..77L}, which is 20-arcmin (40\,cMpc in projection) apart from the SMG structure. There are 91 SMGs with photometric redshifts of $z=4$--5 from S2COSMOS and A$^3$COSMOS \citep{2019ApJ...886...48A,2019ApJS..244...40L}. Some of the SMGs could be at $z=4.6$ and are potentially related with the SMG structure. Future observations will be able to reveal their possible connections and the whole picture of the larger scale structure.

\subsection{Dark matter halo mass}\label{subsec:dmhm}

We estimate the individual dark matter halo masses, $M_\mathrm{h}$, for each of the line emitter SMGs adopting an idealised spherical collapse model \citep[e.g.,][]{2001PhR...349..125B}. This  assumes that the galaxy's measured rotation velocity, $V_{\mathrm{rot}}$, is equal to the circular velocity of the host halo \citep[i.e., isothermal sphere, see also][]{2019A&A...626A..56P} and that the halo is collapsed at the observed redshift. To derive the rotation velocity we first estimate the inclination angle of the [C\,{\sc ii}]-emitting gas disks, $i$, with $i=\cos^{-1}{a_{\mathrm{min}}/a_{\mathrm{maj}}}$, where $a_{\mathrm{min}}$ and $a_{\mathrm{maj}}$ are the deconvolved minor and major axes of [C\,{\sc ii}] emission, respectively (see Table~\ref{tab:lineobs}). The major axes span 4--6\,kpc. We calculate the rotation velocities from the velocity widths with the following relation, $V_{\mathrm{rot}}=1/\gamma\times\mathrm{FWHM}_{\text{[C\,{\sc ii}]}}/\sin{i}$ \citep[$\gamma=2$ ,][]{2019MNRAS.487.3007K}. The range of estimated halo masses of our five SMGs is 2--$8\times10^{12}$\,M$_{\odot}$ within virial radii of 70--110\,kpc. Since it is not clear that the assumption of the flat rotation curve is applicable to our SMGs, we check the consistency with different methods for the halo mass calculation. The mass range of calculated halos corresponds to a bias of $b=8.3\pm1.9$ at $z=4.6$ \citep{2002MNRAS.336..112M}, which is consistent with $b=8.4\pm0.7$ derived from the clustering analysis for $z=3$--$5$ SMGs \citep[e.g.,][]{2019ApJ...886...48A}. In addition, the typical separation of our five SMGs is $\sim10\,\mathrm{cMpc}$ and this is comparable with the predicted clustering length of the halos with mass of $M_h\gtrsim2\times10^{12}$\,M$_{\odot}$ \citep[$\sim$10\,cMpc,][]{2002MNRAS.336..112M}. The main uncertainties in the halo masses come from the inclination measurements of the gas disks. The $\sin{i}$ of AS2COS0001.1, 6.1 and 34.1 are determined with $\sim15\%$ precision. We confirm that the measured inclination and $V_{\mathrm{rot}}$ of AS2COS0001.1 is consistent with those presented in the previous work with higher resolution data within $1\sigma$ uncertainties \citep[$i=51\pm3^{\circ}$ and $V_{\mathrm{rot}}=540\pm90\,\mathrm{km\,s}^{-1}$,][]{2020ApJ...890..171J}. We note that we can only place lower limits on the halo masses for AS2COS0001.2 and 34.2 due to the larger uncertainties of their inclination measurements. Future higher resolution and deeper observations are needed for more precise estimates.

%
%
\begin{figure*}[tp]
\epsscale{1.1}
\centering
\includegraphics[width = 18.2cm,trim = 20.0 10.0 0.0 0.0 cm]{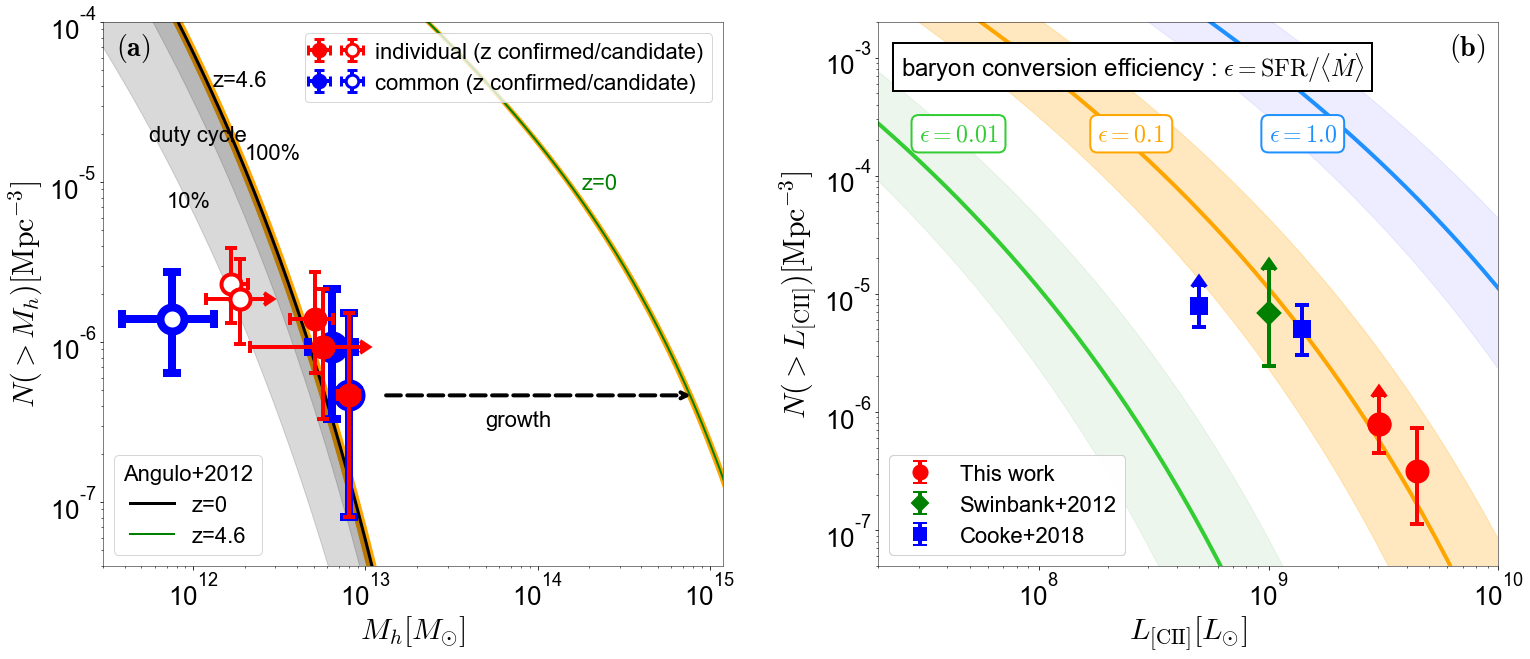}
\caption{({\bf a}) Cumulative number density of dark matter halos for the three [C\,{\sc ii}] emitting SMGs (filled circles) at $z=4.6$ and two candidates (open circles) calculated using a survey volume of $2.2\times10^6\,\mathrm{cMpc}^3$. We plot two cases: where the five SMGs reside in individual halos (red) and where two SMG pairs share common halos (blue).
The black and green curves show the halo mass functions at $z=4.6$ and $z=0$ predicted from a $\Lambda$CDM model \citep{2012MNRAS.426.2046A} with $1$-$\sigma$ uncertainties assuming  cosmic variance \citep[orange shades,][]{2002MNRAS.336..112M,2011ApJ...731..113M}. The gray areas show 50--100\% and 10--50\% of the number densities. We find that the 50--100\% halos with $M_h \geq\,4\,\times 10^{12}$\,M$_{\odot}$ host SMGs at $z=4.6$. In the same survey volume, the expected number of $\sim10^{15}$\,M$_{\odot}$ halos is about unity at $z=0$. ({\bf b})  The [C\,{\sc ii}] cumulative number densities derived from five SMGs and two bright ($S_{870\mu\mathrm{m}}\geq6.2 \mathrm{mJy}$) SMGs in the COSMOS survey area. The value from five SMGs is only a lower limit due to the fact that these are also continuum-selected sources. We show the previous observational constraints at $z\sim4.5$ \citep{2012MNRAS.427.1066S,2018ApJ...861..100C}. Our result constrains the bright end of the luminosity function at $z=4.6$. From the comparison with the model luminosity functions (see text), our constraints imply a baryon conversion efficiency (defined as SFR divided by baryon accretion rate) of $\epsilon \sim 0.1$ at least for the most luminous source.}
\label{fig:hmf+LF}
\end{figure*}

As shown in Figures~\ref{fig:sky} and \ref{fig:mwlimage}, our sample contains two SMG pairs which each could reside in common halos. We estimate their common halo masses from the sky separation and the velocity separation of these pairs assuming an NFW dark-matter profile \citep[Table \ref{tab:lineobs},][]{1997ApJ...490..493N}. We use the concentration parameter of $c=3.5$, as expected for halos of $M_h\sim10^{12}$\,M$_{\odot}$ at $z\sim4.5$ \citep{2014MNRAS.441.3359D}. Since we ignore the line-of-sight separations and the velocities in the transverse direction, the estimated halo masses with NFW profile are lower limits. AS2COS0001.1/1.2 have an angular separation of $3.1''$ ($\sim 21 \,\mathrm{kpc}$) and a redshift offset of $\Delta z=0.010\pm0.002$ ($\Delta v=560\pm50$\,km\,s$^{-1}$ or $\Delta d=1.04\pm 0.20\,\mathrm{pMpc}$). AS2COS0034.1/34.2 have an angular separation of $5.0''$ ($\sim33\,\mathrm{kpc}$) and a redshift offset of $\Delta z=0.006\pm0.002$ ($\Delta v=310\pm60$\,km\,s$^{-1}$ or $\Delta d=0.52\pm0.20\,\mathrm{pMpc}$). The estimated lower limits are $6.4^{+2.3}_{-1.8}\times 10^{12}$\,M$_{\odot}$ for ASCOS0001 and $0.8^{+0.6}_{-0.4}\times 10^{12}$\,M$_{\odot}$ for AS2COS0034, and they are consistent with the halo masses derived from the rotation velocities of the gas disks (Figure~\ref{fig:hmf+LF}(a)).

\section{Discussion}\label{sec:discuss}

\subsection{Nature of  high-redshift SMGs}\label{subsec:hm}

We discuss the nature of $z>4$ bright SMGs in a cosmological context. By comparing with predictions from a $\Lambda$CDM model, we estimate the duty cycle, baryon conversion efficiency and contribution to the cosmic SFRD of the high-redshift SMGs with the full COSMOS survey volume of $2.2\times10^6\,\mathrm{cMpc}^3$.

In Figure~\ref{fig:hmf+LF}(a), we plot the cumulative number density of the five [C\,{\sc ii}] emitting SMGs and the $z=4.6$ halo mass function from \citet{2012MNRAS.426.2046A}. We also show 1-$\sigma$ uncertainty of halo mass function comes from the cosmic variance \citep{2002MNRAS.336..112M,2011ApJ...731..113M}. We find that the 50--100\% halos at $z=4.6$ with a mass of $M_h\gtrsim4\times10^{12}$\,M$_{\odot}$ host SMGs. Such a high duty cycle for bright SMGs could be explained by continuous rapid gas accretion in massive, high-redshift halos or by a series of short duration starbursts triggered via galaxy interactions. We note that the latter interpretation is more consistent with the fact that several of the SMGs may share common halos, where the proximity of the neighboring sources would then also explain the triggering of the intense star-formation within these systems. \citet{2019MNRAS.488.2440M} predicts that mergers are a necessary, but not sufficient, condition to create an SMG: a massive gas reservoir is also needed to power the intense star-formation activity. Therefore a combination of triggering by near-continuous mergers and/or rapid gas fueling by gas accretion might be necessary to produce bright high-redshift bright SMGs with such high duty cycles.

\indent  Owing to the large survey volume of AS2COSMOS, we constrain  $z\sim4.6$ [C\,{\sc ii}] luminosity function brighter than $L_{\text{[C\,{\sc ii}]}}>3\times10^9\,L_{\odot}$ (Figure~\ref{fig:hmf+LF}(b)). Above this luminosity, we have confirmed that completeness correction due to our line search is not necessary for our constraints (see Section \ref{sec:analysis}). Here, we calculate the number densities for two cases, one for the full sample of the five SMGs, and one with the subset of two SMGs with $S_{870\mu\mathrm{m}}\geq6.2\,\mathrm{mJy}$. As our continuum survey is complete to the level of $6.2\,\mathrm{mJy}$, the number density of the full sample of five SMGs gives only a lower limit due to the missing fainter-continuum [C\,{\sc ii}] emitters
in our survey volume. We note that if AS2COS0034.1/34.2 are not [C\,{\sc ii}] emitters, then the number density corresponding to our quoted lower limit should be reduced to 40\% (2/5). With the previous results from \citet{2012MNRAS.427.1066S} and \citet{2018ApJ...861..100C} we find that the bright end of the [C\,{\sc ii}] luminosity function gradually decrease toward $L_{\text{[C\,{\sc ii}]}}\sim10^{10}\,L_{\odot}$.

We compare the [C\,{\sc ii}] luminosity function with model luminosity functions derived from the halo mass function. 
\citet{2009MNRAS.398.1858M} predicts that the mean baryon accretion rate $\langle \dot{M} \rangle$ into a halo as a function of virial mass $M_\mathrm{vir}$ and redshift $z$ is given by the following equation,
\begin{equation}
\begin{split}
    \langle \dot{M} \rangle
    &=f_\mathrm{baryon}\times42\left(\frac{M_\mathrm{vir}}{10^{12}M_{\odot}}\right)^{1.127}\\
    &\times(1+1.17z)\sqrt{\Omega_\mathrm{M}(1+z)^3+\Omega_{\Lambda}}
\end{split}
\end{equation}
Where $f_\mathrm{baryon}$ is the baryon-to-dark matter ratio of $\sim1/6$. The difference of $\langle \dot{M} \rangle$ in the corresponding halo mass range is less than 20\% across the literature \citep{2009MNRAS.398.1858M,2009Natur.457..451D,2015MNRAS.454..637G}. We convert the halo masses to the baryon accretion rates with Equation~1, and calculate SFRs assuming that accreted gas become into stars with baryon conversion efficiencies of $\epsilon=0.01,0.1,1.0$. This baryon conversion efficiency is defined as SFR divided by baryon accretion rate \citep[c.f.,][]{2013ApJ...770...57B,2019MNRAS.488.3143B}. With the SFRs, we calculate $L_{\text{[C\,{\sc ii}]}}$ from the following relation in \citet{2014A&A...568A..62D}.
\begin{equation}
    \log{\rm SFR}=-7.06+1.00\times\log{L_{\text{[C\,{\sc ii}]}}}
\end{equation}
We note that this conversion includes a $1$-$\sigma$ dispersion of $0.27\,\mathrm{dex}$ (shaded areas in Figure~\ref{fig:hmf+LF}(b)). We find that the model luminosity function with a baryon conversion efficiency of $\epsilon\sim0.1$ agrees well with the observed luminosity function at least at the bright end (where our survey is likely complete). The inferred baryon conversion efficiency of $\epsilon\sim0.1$ is similar to that estimated for normal star-forming galaxies with halo masses of $M_h\sim10^{12}$\,M$_{\odot}$, although  at  lower redshifts $z\sim2$, \citep{2013ApJ...770...57B}, which could also have a high duty cycle of $>50\%$ \citep{2005ApJ...631L..13D} and have been claimed to be fed by smooth gas accretion \citep{2009Natur.457..451D}. 
The halos with a mass of $M_h\gtrsim4\times10^{12}$\,M$_{\odot}$ at $z=4.6$ are expected to have a  smooth gas accretion rate of $\gtrsim800$\,M$_{\odot}\,\mathrm{yr}^{-1}$, which is about half of the total baryon accretion rate \citep{2015MNRAS.454..637G}.
Thus the  gas accretion for such halos could be sufficient to achieve the high SFR seen in the SMGs. 
Of course this does not rule out gas accretion through mergers especially given that four of our five targets are potentially close pairs.

%
%
\begin{figure}[tp]
\epsscale{1.15}
\includegraphics[width = 8.5cm,trim = 0.0 10.0 0.0 0.0 cm]{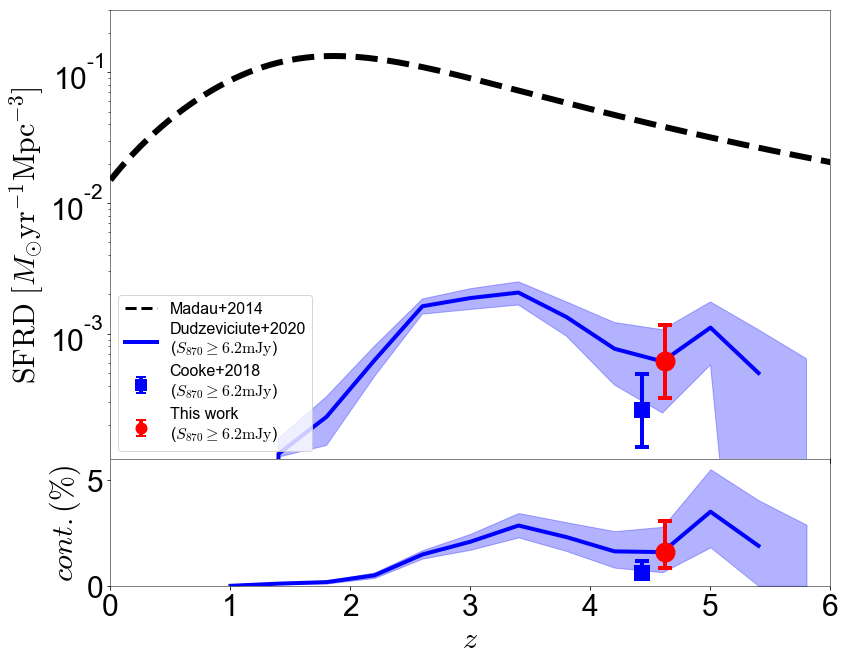}
\caption{The cosmic star-formation rate density \citep[SFRD,][]{2014ARA&A..52..415M} and the contribution from bright ($S_{870\mu\mathrm{m}}\geq6.2\,\mathrm{mJy}$) SMGs. The red filled circle shows the SFRD derived from two brightest SMGs in this survey (AS2COS0001.1/6.1). The blue square and curve show the previous observational constraints from $z\sim4.5$ [C\,{\sc ii}] emitting SMGs and $z=1$--6 SMGs with $S_{870\mu\mathrm{m}}\geq6.2\,\mathrm{mJy}$ \citep{2018ApJ...861..100C,2020MNRAS.494.3828D}. We estimate that the contribution of bright continuum-detected SMGs to the cosmic star-formation rate density is $\simeq2\%$ at $z=4.6$.}
\label{fig:SFRD}
\end{figure}

\indent We estimate the contribution of the bright SMGs at $z=4.6$ to the cosmic star-formation rate density (SFRD). The contribution of dust-obscured galaxies to the cosmic SFRD is less well constrained at $z>4$ \citep[e.g.,][]{2014ARA&A..52..415M,2014MNRAS.438.1267S,2018ApJ...862...77C,2018ApJ...869...71Z,2020MNRAS.494.3828D}. The number density and luminosity of [C\,{\sc ii}] emitters provide robust lower limits on the SFRD at this epoch. We calculate the SFRD of SMGs based on the two bright SMGs for which our survey is complete above its flux level of $S_{870\mu\mathrm{m}}\geq6.2\,\mathrm{mJy}$ across the survey volume (Figure \ref{fig:SFRD}). We convert [C\,{\sc ii}] luminosities to SFR with Equation~2. Here, the error is dominated by the uncertainties from the $L_{\text{[C\,{\sc ii}]}}$--SFR conversion. We confirm that the [C\,{\sc ii}]-based SFRs are consistent with those derived from the dust continuum assuming the relation  in \citet{2020MNRAS.494.3828D}. 

The contribution of the bright continuum-detected SMGs to the total SFRD is $\simeq2~\%$ at $z=4.6$. This is comparable with the contribution of dust continuum selected SMGs presented in previous works when we use the same flux cutoff  \citep[$S_{870\mu\mathrm{m}}\geq6.2\,\mathrm{mJy}$,][]{2018ApJ...861..100C,2020MNRAS.494.3828D}. In contrast, \citet{2018ApJ...869...71Z} claimed that the contribution of dust-obscured galaxies at $z>4$ is $\sim35$--85\% based on an ALMA 3-mm survey. The apparent lower contribution of the bright SMGs may come from our high flux cutoff. The flux cutoff of $S_{870\mu\mathrm{m}}\geq6.2\,\mathrm{mJy}$ corresponds to $\mathrm{SFR}=330\,\mathrm{M}_{\odot}\,\mathrm{yr}^{-1}$ \citep{2020MNRAS.494.3828D}, which is $\sim4$ times higher than the flux limit in \citet[][assuming modified black body with $T_d=35$\,K and $\beta=1.8$]{2018ApJ...869...71Z}. In addition, as extensively discussed by \citet{2018ApJ...869...71Z}, their sample may also be biased due to the survey including sources associated with the primary targets in the archival fields or due to synchrotron contributions to their 3-mm fluxes. We find that the cosmic SFRD can be reproduced when we integrate our model of [C\,{\sc ii}] luminosity function with $\epsilon\sim0.1$ down to $L_{\text{[C\,{\sc ii}]}}=10^{7.5}L_{\odot}$ and convert the [C\,{\sc ii}] luminosity density to SFRD using the Equation~2. This indicates that we need at least $\sim100$ times deeper observations to place a meaningful constraint to the cosmic SFRD from [C\,{\sc ii}] luminosity function.

\subsection{SMG large-scale structure and the descendant}\label{subsec:evo}

The rough  alignment of the five SMGs apparent in Figure~\ref{fig:sky} may suggest that simultaneous  star-formation activity is occuring in massive halos along the cosmic web.
Numerical simulations predict that more than 90\% of dark matter is in the form of un-virialized filamentary large scale structure at $z\sim5$ \citep{2016MNRAS.457.3024H}. 

The scale of the SMG-traced structure is comparable with expected size of protoclusters at $z=4.6$ \citep{2013ApJ...779..127C}. Numerical simulations suggest that most halos with a mass of $M_h\gtrsim2\times10^{12}$\,M$_{\odot}$ at $z=4.6$ evolve into a $M_h\gtrsim10^{15}$\,M$_{\odot}$ halo by $z=0$ \citep[e.g.,][]{2009ApJ...707..354Z}, which corresponds to the present day's most massive cluster. As shown in Figure \ref{fig:hmf+LF}(a), the expected number of $M_h\sim10^{15}$\,M$_{\odot}$ halos is about unity in our survey volume at $z=0$. The present-day $\sim10^{15}$\,M$_{\odot}$ halos are expected to have a maximum baryon conversion efficiency of $\epsilon\sim0.1$ at $z\sim4$--$7$ \citep{2013ApJ...770...57B}, which is consistent with our findings (see Section \ref{subsec:hm}). We suggest, therefore, that we are witnessing  efficient, synchronized star formation in massive galaxies, driven by mergers and potentially smooth gas accretion, within a  massive protocluster environment at $z\sim4.6$ when the Universe was only 10\% of its current age.

\section{Summary} \label{sec:summary}

\indent We detected emission lines from six SMGs based on ALMA S2COSMOS Band~7 observations of 184 luminous submillimeter sources with $S_{850\mu{\rm m}}\,\geq\,6.2\,\mathrm{mJy}$ in the full $1.6~\mathrm{deg}^2$ COSMOS field \citep{2020MNRAS.495.3409S}. Among these, four SMGs have been confirmed to be [C\,{\sc ii}](157.74\,$\mu$m) emitters at $z=4.60$--$4.64$ by independent detections of $^{12}$CO($J$=5--4) emission. The remaining two SMGs are also likely to be [C\,{\sc ii}] emitters at $z=4.62$ from their line equivalent widths and multi-wavelength spectral energy distributions.  Four of the six SMGs are near-infrared blank SMGs. After excluding one SMG whose emission line is falling at the edge of spectral window, all line emitting SMGs show clear velocity gradients along the major axes of their [C\,{\sc ii}] emission, consistent with gas disks with rotation velocities of 330--550\,km\,s$^{-1}$. We estimate that they have individual dark matter halo masses of $2$--$8\times 10^{12}$\,M$_{\odot}$, a 50--100\% duty cycle, and a baryon conversion efficiency (SFR relative to baryon accretion rate) of $\epsilon \sim 0.1$ and contribute $\simeq2\%$ to the total SFRD at this redshift. Five SMGs are concentrated within a $16\times4\times12\,\mathrm{cMpc}^3$ region and likely evolve into members of a massive ($\sim10^{15}$\,M$_{\odot}$) cluster at $z\sim 0$. Our work demonstrates that the combination of wide-field single dish survey and ALMA follow-up is a powerful method to investigate the rapid formation process of massive galaxies within the cosmic web at $z>4$. 


%
%
%
\acknowledgments

We would like to thank the anonymous referee for a thoughtful and constructive report which improved the content and clarity of this paper. We also thank to C.\ Casey, H.\ Nagai and T.\ Michiyama for the useful discussion and information. YM, HU and HI acknowledge support from JSPS KAKENHI Grant (17H04831, 17KK0098, 19H00697, 17K14252, 20H01953 and 19K23462). IRS, JMS, UD, AMS \& JEB acknowledge support from STFC(ST/T000244/1). CCC acknowledge support from the Ministry of Science and Technology of Taiwan (MOST 109-2112-M-001-016-MY3). This paper makes use of the following ALMA data: ADS/JAO.ALMA \#2013.1.00034.S, \#2015.1.00137.S, \#2015.1.00568.S, \#2015.1.01074.S, \#2016.1.01604.S, \#2016.1.00478.S, \#2016.1.00463.S. ALMA is a partnership of ESO (representing its member states), NSF (USA) and NINS (Japan), together with NRC (Canada) and NSC and ASIAA (Taiwan), in cooperation with the Republic of Chile. The Joint ALMA Observatory is operated by ESO, AUI/NRAO and NAOJ. Data analysis were carried out on common use data analysis computer system at ADC/NAOJ.

\bibliography{draft1.bib}{}
\bibliographystyle{apj.bst}

\end{document}